# Inverse Photoelectrochemical Cell

Qiang Yu and Chuanbao Cao*

Research Center of Materials Science, Beijing Institute of Technology, Beijing 100081, China

*Email: yuqiang@bit.edu.cn, cbcao@bit.edu.cn

**Abstract:** The splitting of water with sunlight using photoelectrochemical cell (PEC) to produce hydrogen is a promising avenue for sustainable energy production. The greatest virtue of PEC is that it uses sunlight as the only source to split water, but its efficiencies are still quite low due to poor performances of the available materials (such as $SrTiO_3$). Consequently, another way of PEC research has been developed. By simultaneously using sunlight and electricity as energy source, PEC can get a larger current at a lower voltage, i.e., hydrogen can be made under the voltage below 1.23V, the minimum voltage required by electrolysis of water. But so far the efficiencies of the mainstream materials (such as $TiO_2$) remain low. Where is the future development direction of PEC? Here we propose a new PEC model. Its operating principle is quite the opposite of the aforesaid conventional PEC, that is, the previous photoanode/photocathode is converted into the present photocathode/photoanode. It can also obtain high current under low voltage, even near zero voltage in extreme conditions. A basic single configuration and an improved n-p configuration were designed. We reselected materials for photoelectrodes, and confirmed its feasibility successfully. Two preliminary results were obtained from two contrasts: whether compared with water electrolysis or conventional PEC water splitting, the inverse PEC showed promising superiority.

The splitting of water with sunlight using photoelectrochemical cell (PEC) to produce hydrogen is one of the most sustainable forms of energy production, since both water and sunlight are vastly abundant.[1] A great deal of research has been done since the discovery of PEC water splitting in 1972.[2] PEC has become a major topic of renewable energy research, with over 3000 papers published in the last five years. A typical PEC consists of two electrodes immersed in an electrolyte solution, namely anode and cathode. Schematic diagram is shown in Figure 1a, taking an n-type semiconductor photoelectrode and a metal counter electrode (commonly platinum) as an example. Contact between two electrodes immersed in the electrolyte (no illumination) results in charge transfer until the work functions of both electrodes equilibrate (Figure 1-a0). If anode is n-type semiconductor and cathode is metal, under light illumination, the excited electrons move to cathode where water reduction occurs, and the generated holes move towards anode/electrolyte interface where water oxidation takes place. The most ideal model of PEC is that the cell only uses sunlight as the energy source to split water, but an important prerequisite is required. For n-type semiconductor photoanode, the Fermi level (or flat band potential, $V_{FB}$) should be located above the ($H^+/H_2$) water reduction potential (Figure 1-a1). Unfortunately, most of photoanodes cannot meet this requirement, and several available semiconductors satisfying the criteria (such as $SrTiO_3$) usually have large bandgap energies that result in low optical absorption and hence low photoconversion efficiencies.[3] The self-biased n-p tandem PEC which relaxes the criteria even still has quite low efficiencies.[4,5] Consequently, another way of PEC research has been developed. Using sunlight and electricity simultaneously as energy sources, PEC can split water under the voltage below 1.23V, the theoretical minimum voltage required by electrolysis of water. That is to say, this kind of PEC can get a larger current at a lower voltage. Because of the external bias, $H_2$ can be produced even when the flatband potential of semiconductor is below the $H^+/H_2$ level (Figure 1-a2). Therefore it seems more accurate to call it "photo-assistant electrolysis". Today the majority of PEC researches belong to this category. The widely tested photoelectrode materials include n-$TiO_2$,[6] n-$Fe_2O_3$,[7] n-ZnO,[8] n-$WO_3$,[9] p-Si,[10.11] p-$Cu_2O$,[12] p-GaP,[13] p-$GaInP_2$,[14] p-$CuGaSe_2$,[15] etc. Although the performances of these materials have been improved significantly after years of exploration, the conversion efficiencies or stabilities are still far from practical application due to various reasons. For example, n-$TiO_2$ shows limited photocurrent typically less than 1 mA/cm$^2$ because of its large bandgap.[16] Thus it can be seen that the progress in improving PEC performances, in both research directions of PEC above, is slow. In view of this situation, it seems urgent to ask a question: "Where next?" How much efficiencies of PEC could we obtain? Can we identify promising routes that will take us to high PEC efficiencies which can meet the requirement for the commercial application?

Here we propose a new PEC mechanism as an alternative model of PEC researches. By coincidence, the new operating principle is quite the opposite of the above mentioned conventional PEC's. Simply put, the previous photoanode is converted into present photocathode and so as photocathode is converted into photoanode. We name it "inverse PEC" (IPEC). Similar to the second conventional PEC mentioned above, the effect that the new PEC model tries to realize is also to obtain high current under low voltage, i.e., producing hydrogen under the voltage lower than 1.23 V. Can "n-type cathode" or "p-type anode" work? The energy diagram of inverse PEC is shown in Figure 1b, also taking an n-type semiconductor photoelectrode and a metal counter electrode as an example. In the conventional PEC, n-type semiconductor acts as photoanode, anodic bias provides overvoltage to sustain the current flow, making the photo-excited electrons move to the metal cathode where $H_2$ reduction occurs (Figure 1a). In the case of our new model, cathodic bias is applied to the n-type photoelectrode, thus the electrons will transfer from semiconductor to electrolyte if the $V_{FB}$ of semiconductor locate above the ($H^+/H_2$) level, then hydrogen evolution will occur at this semiconductor electrode (Figure 1-b1). Here n-type semiconductor acts as cathode. Because electron in n-semiconductor is majority carrier, it can be predicted that the electrochemical behavior of semiconductor electrode will be similar to that of metal electrode, and large current is expected to be obtained. Then what about "n-type photocathode"? Under the illumination, the semiconductor surface Fermi level is elevated. We define this value of potential change as Vn (that is photovoltage). Hence, when the same voltage is reached between semiconductor cathode and metal anode by applying cathodic bias to the semiconductor cathode, the external voltage of Vn can be less applied to cell under light illumination than in the dark. So n-type photocathode has the effect of reducing external applied voltage, and it is the photovoltage Vn that enables the inverse PEC to get large current under low voltage.

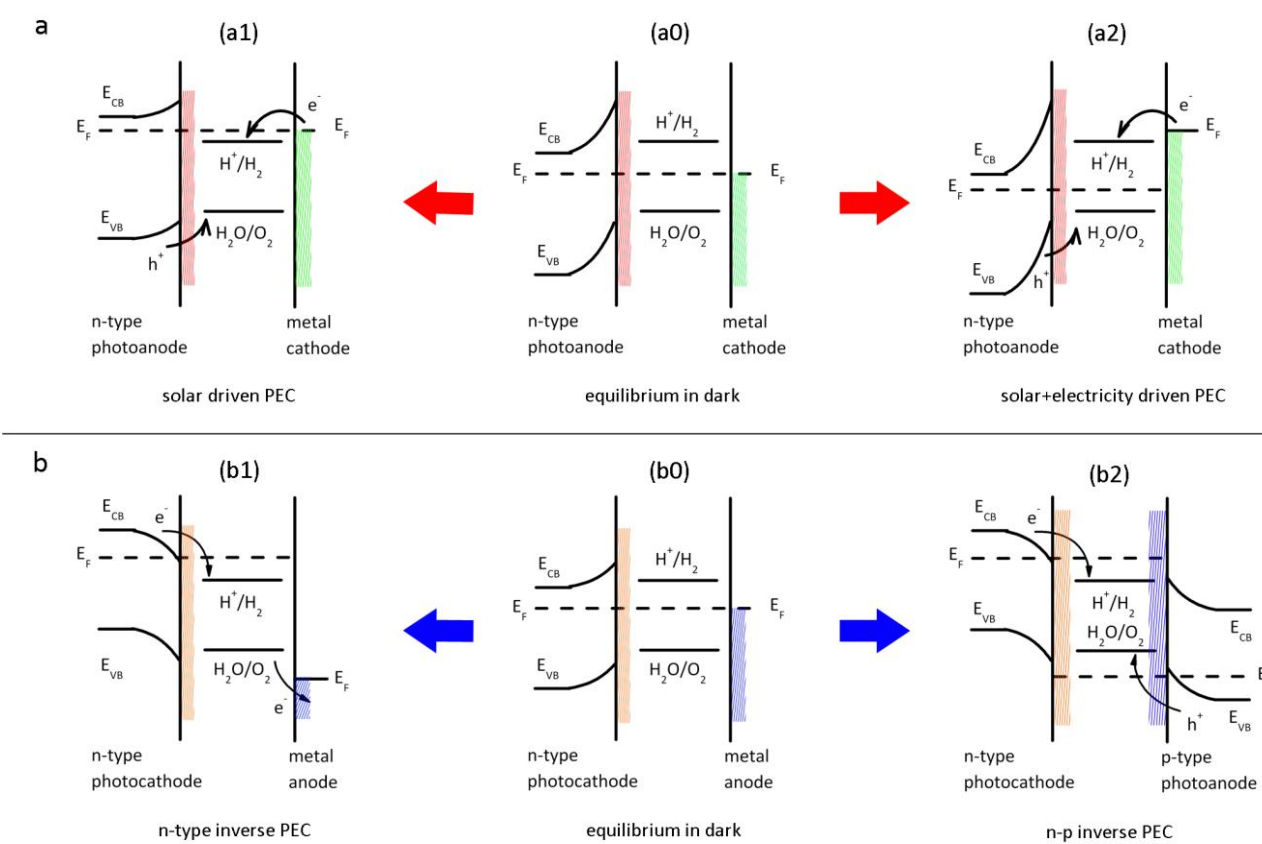

***Figure 1.*** Principle of operation of conventional PEC (a) and new inverse PEC (b) with the example of n-type semiconductor. (a0)/(b0) equilibrium of two electrodes immersed in electrolyte in dark; under illumination, one conventional PEC works driven by only solar energy (a1)and another conventional PEC works driven by solar and electricity simultaneously (a2). Under light illumination and external voltage, the inverse PEC consisting of a single photoelectrode (b1) and two photoelectrodes (b2).

While this new operating mechanism is simple, what we really want is the inverse PEC can possibly get better performance than the conventional PEC. So some comparisons are needed. Here we choose two devices as the comparison objects. The first is water electrolysis with metal electrodes (here is Pt). Electrolysis can generate large current, but it needs applied external voltage larger than 1.23 V. The second is conventional PEC water splitting. The applied voltage it needed could be lower than 1.23 V, but the current it achieved now is relatively small. Can the new inverse PEC surpass the above two model? We speculate that, as indicated in Figure 1-b1, the inverse PEC will decrease applied voltage in comparison with the Pt electrode electrolysis, but the external voltage remains required inevitably and could possibly be still larger than the value applied in the conventional PEC; the current of the inverse PEC should exceed the conventional PEC, but seems to remain lagging behind the electrolysis. Consequently, the key to winning is to reduce voltage as much as possible. Therefore, based on the single n-type photocathode cell above, we further design an improved n-p configuration for inverse PEC, i.e., combining n-semiconductor photocathode with p-semiconductor photoanode. As shown in Figure 1-b2, we apply anodic bias to p-semiconductor, making it act as anode. Under the illumination, the semiconductor Fermi potential moves in the anodic direction. The value of this potential change is defined as Vp. So the saved external voltage of the n-p inverse PEC will increase to Vn+Vp. Going even further, if it exists that the conduction band of some appropriate n-type materials and the valance band of the p-type semiconductors span the water redox reactions, i.e., Vn+Vp > 1.23 V, the applied external voltage needed to realize water splitting will be further reduced, even possibly as low as near zero in theory. Hence the n-p structure is the real promising inverse PEC model which can reduce bias as much as possible while simultaneously maintaining high current. In other words, the n-p inverse PEC combines the virtues of water electrolysis and conventional PEC and is expected to be a competitive alternative model of the conventional PEC. In addition, if the inverse PEC can consume hardly electricity, most of hydrogen productions come from solar energy, the conversion efficiency would be expected to increase significantly. From these points, the inverse PEC is not only an alternative model of PEC researches, but also a superior model better than conventional PEC.

Below we set up the experiment to test this new idea. Because the inverse PEC is a different configuration, the present common photoelectrode materials might be no longer applicable, so we have to reselect materials for photoelectrdes. We think that the selection of photoelectrode materials should comply with following principles: the conduction band edge of n-type semiconductor should lie at a position more negative (NHE as reference) relative to the reduction potential of water and the valence band edge of p-type material more positive compared to the oxidation potential. In many traditional favourable n-type photoanode materials, conduction band edge is located more positive relative to the ($H^+/H_2$) potential,[17] such as, n-$Fe_2O_3$, n-$WO_3$ and n-ZnO (Figure 2). So these materials are not good options for IPEC. n-Si is not a proper photoanode because its valence band edge potential is not sufficiently positive for water oxidation. But n-Si has conduction band edge potential that is positive of the ($H^+/H_2$) potential to drive the hydrogen evolution reaction. Thus we choose n-Si as the photocathode of the inverse PEC. For p-type photoanodes, many common p-type photocathode materials of conventional PEC do not satisfy the criteria due to insufficiently positive valence band (p-Si, p-$Cu_2O$, etc.). Here we use p-CuO that has more positive valence band as photoanode, though it is not an eligible traditional photocathode material. The experiments were carried out to test the performances of inverse PEC devices comprised of p-CuO photoanode and n-Si photocathode under the illumination/in the dark, and compare it with water electrolysis (use Pt electrode).

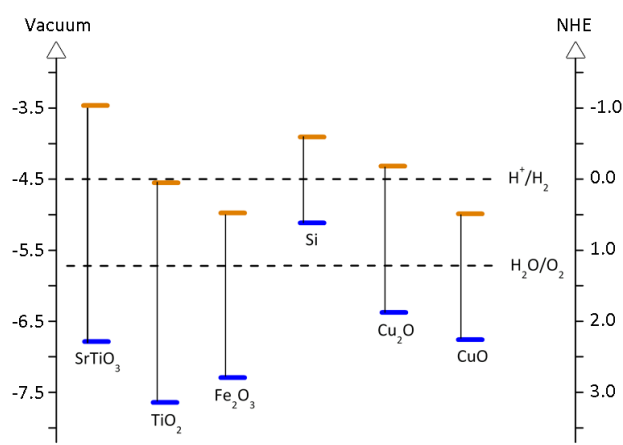

*Figure 2. Band edge positions of several semiconductors used in PEC cells.*

**Results and discussion**

Photoelectrochemical experiments were first carried out in the two-electrode systems. Figure 3a shows representative I-V curve of the electrolytic cell comprised of Pt anode and Pt cathode under our experiment conditions. The electrolyte was 0.5 M KOH (pH=13) and simulated sunlight (100 mW/$cm^2$, AM 1.5) was used as the light source. In dark, under the external 4 V voltage, Pt electrode (0.2 $cm^2$) achieved a large current density of 200 mA/$cm^2$, which started nearly at 2 V (~2 V). The threshold voltage is higher than the theoretical value of 1.23 V, indicating that even with Pt electrodes large overpotentials are needed for water splitting. The conventional PEC could decrease the threshold voltage, but the photocurrents achieved were quite low. In contrast, the inverse PEC here may obtain large currents and reduced voltages simultaneously. Figure 3b shows the I-V curves of the inverse PEC consisted of n-Si photocathode and Pt anode under the applied external voltage of 0-4 V. In the dark, the cell yielded large current of 70 mA/$cm^2$ and current onset of ~2 V; under light illumination, the current increased to 120 mA/$cm^2$ and the threshold voltage decreased to ~1.8 V. Large amounts of bubbles generated from the anode and cathode were observed, and the growth and release of the big bubbles on the electrode surfaces was the reason why the I-V curves were zigzag in the large current regions. These results successfully demonstrated two predictions: (1) The large currents demonstrate it is feasible to use n-type semiconductor as cathode. The semiconductor now is more like a metal, to be precise, it is a majority-carrier (electrons) device. Under illumination some electrons could be excited to jump into the conduction band in semiconductors, but the amount of the photogenerated carriers is far fewer than the number of electrons in the conduction band, so the photogenerated electrons should not make great contributions to the large currents. (2) The reduced threshold voltage confirms the effects of the inverse PEC. That is exactly the result of the negative shift of the Fermi level of the n-type semiconductor under illumination. In our testing we found that the potential of n-Si was -0.45 V (vs SCE) in dark (tested in three-electrode system), and it shifted to -0.60 V under the light, with the variation of -0.15 V. Correspondingly, in I-V curves the threshold voltage of the cell similarly shifted ~0.2 V. Hence, the single photocathode inverse PEC can indeed reduce external bias while simultaneously maintaining high current. The role of the n-Si photocathode is reflected in the extra voltage it supplied under light illumination which can save part of external applied voltage. If we find some appropriate materials having more negative conduction band, the generated photovoltage would become greater.

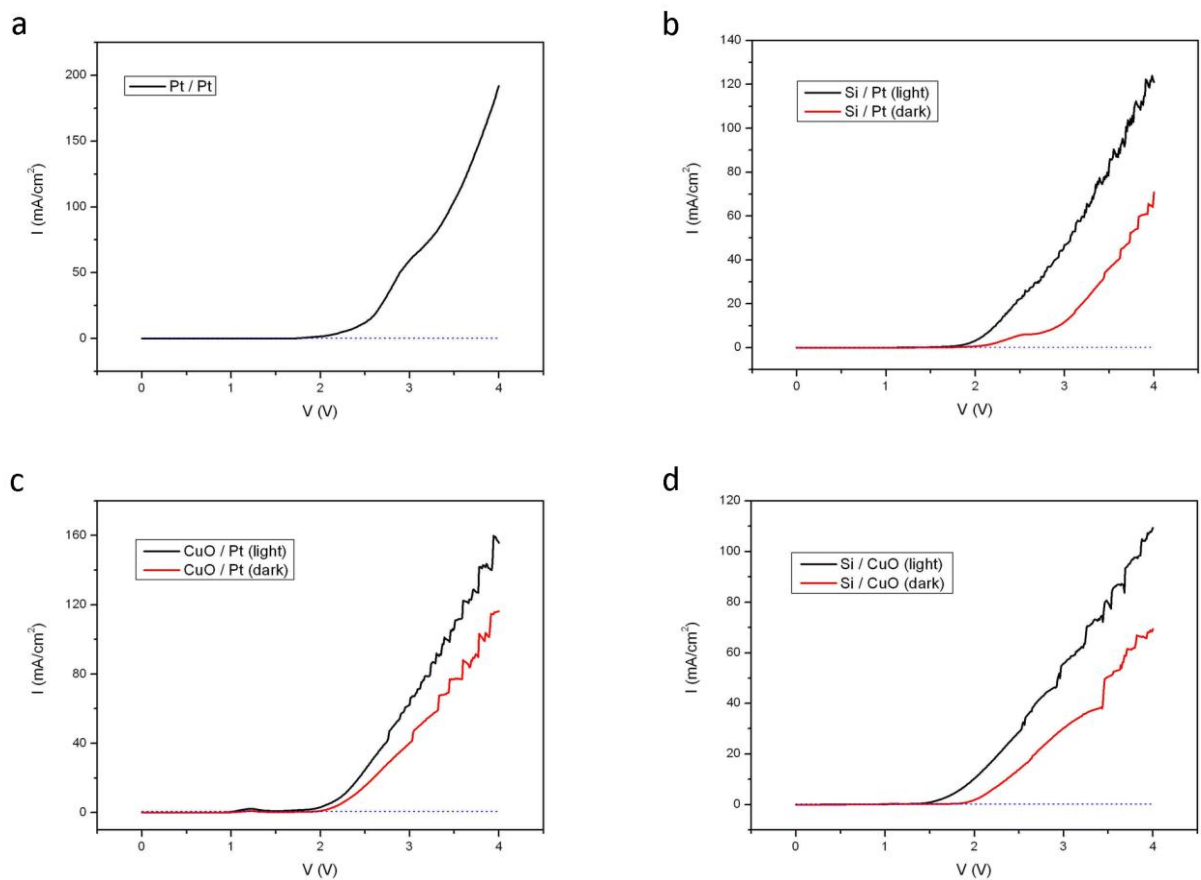

*Figure 3.* Current-voltage curves of (a) electrolysis using two Pt electrodes in dark, (b) the inverse PEC cell consisting of an n-Si photocathode and a Pt anode, (c) a CuO photoanode and a Pt cathode and (d) an n-Si photocathode and a CuO photoanode, tested in the two-electrode systems in 0.5 M KOH electrolyte (pH=13) under illumination / in dark. Simulated sunlight (100 mW/cm2, AM 1.5) was used as the light source. The direction of applied external voltage was that anode was positive and cathode was negative.

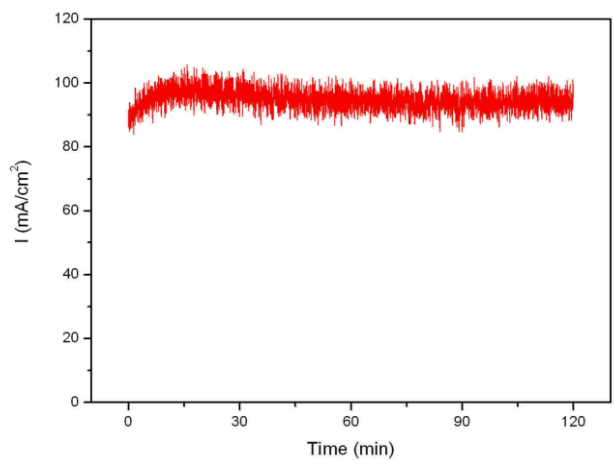

*Figure 4.* I-t curve recorded from the n-p inverse PEC consisting of an n-Si photocathode and a CuO photoanode for a 2 h duration of operation at applied voltage of 4 V under illumination.

The p-CuO photoanode had I-V behavior similar to the n-Si photocathode in the inverse PEC. As shown in Figure 3c, in dark, the cell exhibited current of 120 mA/cm$^2$ and current onset of ~2 V; in light, the current increased to 150 mA/cm$^2$ and the threshold voltage decreased to ~1.9 V. So this result demonstrated again the feasibility of p-type photoanode. But the potential of p-CuO appeared to be not sufficiently positive. The potential of the p-CuO was recorded to be 0 V (vs SCE) in dark and +0.10 V in light. Figure S2 (Supplementary Information) displays the differential capacitance vs potential data in the form of 1/C$^2$ vs E (Mott-Schottky plot) collected from the p-CuO photoanodes and the n-Si photocathodes. This is a classical method of determining the flat band potential of a semiconductor electrode.[18] $V_{FB}$ was determined from the extrapolation of the linear portion of 1/C$^2$ vs. E to the x-axis intercepts in Mott-Schottky plots. The n-Si and p-CuO had $V_{FB}$ of -0.60 V and +0.35 V, respectively. In our testing, the potential of the n-Si under illumination was near this inferred flat band potential, indicating the Fermi level is just below the conduction band edge. But the potential of the p-CuO was far away from +0.35 V, probably due to the large resistivity or other reasons and further investigations are needed. Next, we tested the n-p inverse PEC which connected, in series, an n-Si photocathode and a p-CuO photoanode. This configuration can add the photovoltages from two photoelectrodes together. As shown in Figure 3d, the threshold voltage was further reduced to 1.6 V. These experiments showed successfully the feasibility of our newly designed inverse PEC, though there was not an apparent significant decrease as much as expected in threshold voltage due to the insufficiently positive potential of p-CuO. Here the total photovoltage of the n-p inverse PEC just reached 0.70 V, less than 1.23 V. If the proper materials which can generate photovoltages greater than 1.23 V or larger could be found in the future, the threshold voltage to split water should be further decreased, might even be close to zero, in theory. Consequently, the improved n-p inverse PEC shows a greater capacity to reduce the external voltage and also shows the potential to further decrease the bias. Moreover, it's worth noting that the long-term operation test of the n-p inverse PEC cell showed reasonably steady current (Figure 4), which demonstrated good chemical stability and corrosion resistance of both n-Si photocathode and p-CuO photoanode. Maybe the selection of photoelectrodes made contributions to this effect. For example, n-Si could be subject to oxidization in aqueous solution, but when it is used as photocathode the cathodic current may, to some extent, protect the surface of the semiconductor from oxidation.

The above comparisons between the water electrolysis and the inverse PEC successfully demonstrated the capacity of the inverse PEC to reduce the voltage. The inverse PEC also has the potential to possibly further decrease the voltage. So at present compared with the conventional PEC which can also split water under the voltage below 1.23 V, is the inverse PEC better? It can be speculated that the inverse PEC will exceed the conventional PEC started at one bias below 1.23 V. Therefore, we also compared the inverse PEC with the conventional PEC in common used three-electrode systems. Before some excellent electrode materials are found, we first use the existing materials to implement preliminary tests. Here we selected n-$TiO_2$ owing to its ability to act as both photoanode in conventional PEC and photocathode in inverse PEC, though it is not the optimization for inverse PEC due to its insufficiently negative conduction band. As shown in Figure 5a, the conventional PEC consisting of an n-$TiO_2$ photoanode and a Pt cathode obtained the current of 0.35 mA/cm$^2$ under light illumination until the external anodic bias exceeded 1.5 V, from which water electrolysis was started. It showed that the conventional PEC can get photocurrent under the voltage below 1.23 V, though the current remained relatively small (for example, 0.3 mA/cm$^2$ at 0.2 V). Figure 5b shows the I-V curves of the inverse PEC composed of an n-$TiO_2$ photocathode and a Pt anode under applied cathodic bias of 0-1.5 V. In light, the cell showed large current of 40 mA/cm$^2$ and current onset of ~0.8 V. From the magnified image (Figure 5c) it can be seen more clearly that over bias of 0.8 V the inverse PEC showed sharply increased current much larger than conventional PEC. This result demonstrated our prediction that the inverse PEC can exceed the conventional PEC in the range of 0.8-1.5 V. Surprisingly, below 0.8 V the inverse PEC also yielded larger photocurrent than the conventional PEC (for example, 0.4 mA/cm$^2$ at 0.2 V). That is to say, whether in the range of 0.8 – 1.5 V or 0 – 0.8 V, the inverse PEC yielded larger currents than the conventional PEC, greatly or slightly. The original purpose of this test with the example of commonly used traditional $TiO_2$ photoelectrode material is to show that the inverse PEC definitely has the possibility to surpass the conventional PEC made from the same material. Yet now in term of $TiO_2$, we could almost say that the inverse PEC had a better performance than the conventional PEC. $TiO_2$ is not an optimal choice for the inverse PEC and has decreased stability in the inverse PEC. The threshold voltage from which the inverse PEC can real far exceed the conventional PEC (here was 0.8 V) was still large. In future, more appropriate materials are required to have lower threshold voltage.

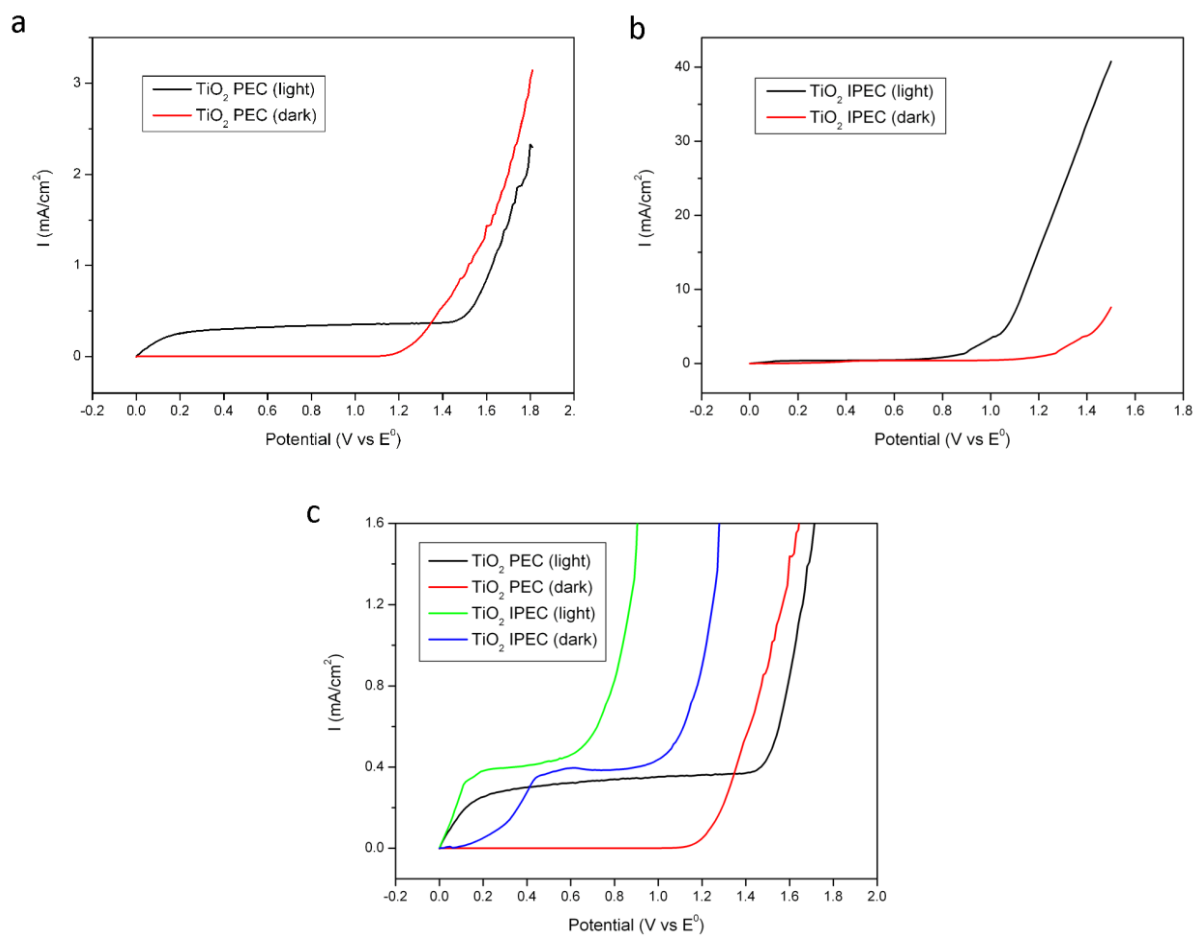

*Figure 5.* I-V curves of (a) a conventional PEC consisting of n-$TiO_2$ photoanode and Pt cathode and (b) an inverse PEC consisting of n-$TiO_2$ photocathode and Pt anode. (c) The magnified image of the comparison of I-V curves between (a) and (b) in the range of low currents. The cells were tested in 0.5 M KOH electrolyte using three-electrode systems with n-$TiO_2$ photoelectrode as the working electrode and SCE and Pt as the reference and counter electrodes, respectively. Simulated sunlight (100 mW/cm$^2$, AM 1.5) was used as the light source. For better comparison, absolute values of the bias and current (anodic and cathodic) were taken to plot in one diagram. $E^0$ was the initial potential of $TiO_2$ in PEC cell and was defined as 0 V.

In summary, we proposed a new PEC mechanism (inverse PEC) as an alternative model. A basic single structure and an improved n-p inverse PEC structure were designed. The inverse PEC has the capacity to combine the virtues of water electrolysis and conventional PEC, i.e., reducing external voltage (below 1.23 V) while maintaining high current simultaneously. And the inverse PEC also shows the potential of further decreasing the bias, possibly as low as near zero in theory. Therefore the inverse PEC can be expected to be a competitive alternative model of the conventional PEC. We established selection principles of IPEC photoelectrodes and reselected materials. By taking n-Si, p-CuO and n-TiO$_2$ as examples, we confirmed successfully the feasibility of the inverse PEC. Two comparison experiments were performed and the results were consistent with our expectations: compared with water electrolysis, IPEC showed reduced threshold voltages and considerable currents; compared with the conventional PEC, IPEC can far exceed the conventional PEC starting at one bias below 1.23 V. These materials only satisfied basic requirements and their performances were far from perfect results (using none of catalysts was another reason). But we believe more optimal materials will be found soon. The new inverse PEC displays a bright future: if one day, PEC water splitting work efficiently under extraordinarily low voltages, the application of solar-to-hydrogen renewable energy supply will really enter into millions of households.

**Methods**

**Photoelectrodes Preparation:**

**Si:** Single-side polished n-type Si(100) wafers (<0.01 Ωcm resistivity, 450µm thick) were diced into approximately 2 cm × 2 cm square pieces. Silicon samples were cleaned with acetone, methanol, and water sequentially. The samples were subsequently oxidized in H$_2$O$_2$/H$_2$SO$_4$ 1:3 at 90 ℃ for 10 min to remove heavy metals and organic species, etched for 1-2 min in buffered HF to remove native oxide, then rinsed with deionized water, and dried under a stream of N$_2$. Ohmic contacts were made by immediately rubbing Ga-In eutectic on the oxide-free, unpolished back sides of the n-Si chips, followed by attachment of a coiled tin-copper wire using conductive silver paint. The exposed backside, edges, and some part of the front of the samples were sealed with nonconductive hysol epoxy, leaving a typical active area of 0.25cm$^2$ of the front-side Si surface exposed to the electrolyte.

**CuO:** Copper oxide thin films were obtained by annealing the metallic Cu foils in a furnace. Prior to oxidation, industrial-grade copper sheets (0.25 mm thickness and 99.99% purity) were cleaned in 30% nitric acid for 20 s and cut into standard sizes of 4 cm × 2 cm. The samples were then repeatedly rinsed in deionised water and dried with tissue paper. The oxidation was carried out at 500 ℃ in open air for 2 h. The electric contacts and sealing of the CuO samples were prepared following the same treatments for the Si samples, except affixing the copper wire directly to the bare portion of Cu substrate under CuO with Ag paint.

**TiO$_2$:** A thin film of TiO$_2$ was sputtered onto a fluorine-tin oxide coated glass substrate (FTO, 30 Ω per square) at room temperature in an argon gas atmosphere using a titania ceramic target. The working pressure, RF power and deposition time were held constant at 7 mTorr, 150 W and 4 h, respectively. After deposition, the samples were annealed at 450 ℃ for 2 h in air. The n-TiO$_2$ photocathode was fabricated similar to the n-Si photoanode, also with an effective working area of 0.25 cm$^2$, except securing the copper wire directly onto the bare portion of FTO substrate with Ag paint.

**Photoelectrochemical Measurements:**

PEC measurements were performed in both standard three-electrode configuration and two-electrode system. In the two-electrode system, external voltage was applied between anode and cathode. Conventional current flow direction is used, i.e., anode is positive electrode (applying anodic voltage) and cathode is negative electrode (applying cathodic voltage). Here n-Si electrode served as photocathode or p-CuO electrode served as photoanode, the counter electrode was Pt sheet. In the three-electrode configuration, the photoelectrode acted as working electrode, with a Pt sheet as counter electrode and a saturated calomel electrode (SCE) as reference electrode (+0.241V versus NHE, Normal Hydrogen Electrode). The electrolyte solutions are 0.5M KOH (pH=13) aqueous solution.

Simulated sunlight (100 mWcm$^{-2}$, AM 1.5, Oriel 91160) was used as the light source. The illumination intensity was calibrated from outside of the PEC cell. All PEC and impedance measurements were carried out on a ZAHNER IM6e potentiostat. Linear sweep voltammagrams (I-V) were measured at a scan rate of 5mV/s. The impedance measurements were performed using the same three-electrode cell configuration and instrument as the photoelectrochemical measurements. The impedance measurements (Mott-Schottky) were performed with 5 mV amplitude modulation on a reverse direct current bias in the dark; the frequency range was between 5 and 10 kHz. I-V measurements in the dark were repeated before and after each applied potential of impedance measurements to ensure that no irreversible changes occurred to the photoelectrode under investigation. The electrochemical data were analyzed and reported as-collected, without correction for any solution resistance or concentration overpotential losses.

**References**

[1] M. G. Walter, E. L. Warren, J. R. McKone, S. W. Boettcher, Q. X. Mi, E. A. Santori, N. S. Lewis, *Chem. Rev.* **2010**, *110*, 6446.
[2] A. Fujishima, K. Honda, *Nature* **1972**, *238*, 37.
[3] I. E. Paulauskas, J. E. Katz, G. E. Jellison, Jr., N. S. Lewis, L. A. Boatner, *Thin Solid Films* **2008**, *516*, 8175.
[4] A. J. Nozik, *Appl. Phys. Lett.* **1976**, *29*, 150.
[5] H. L. Wang, T. Deutsch, J. A. Turner, *J. Electrochem. Soc.* **2008**, *155*, F91.
[6] J. Nowotny, T. Bak, M. K. Nowotny, L. R. Sheppard, *Int. J. Hydrogen Energy* **2007**, *32*, 2609.
[7] K. Sivula, F. Le Formal, M. Graetzel, *Chemsuschem* **2011**, *4*, 432.
[8] K. S. Ahn, S. Shet, T. Deutsch, C. S. Jiang, Y. F. Yan, M. Al-Jassim, J. Turner, *J. Power Sources* **2008**, *176*, 387.
[9] V. Chakrapani, J. Thangala, M. K. Sunkara, *Int. J. Hydrogen Energy* **2009**, *34*, 9050.
[10] S. W. Boettcher, E. L. Warren, M. C. Putnam, E. A. Santori, D. Turner-Evans, M. D. Kelzenberg, M. G. Walter, J. R. McKone, B. S. Brunschwig, H. A. Atwater, N. S. Lewis, *J. Am. Chem. Soc.* **2011**, *133*, 1216.
[11] S. W. Boettcher, J. M. Spurgeon, M. C. Putnam, E. L. Warren, D. B. Turner-Evans, M. D. Kelzenberg, J. R. Maiolo, H. A. Atwater, N. S. Lewis, *Science* **2010**, *327*, 185.
[12] A. Paracchino, V. Laporte, K. Sivula, M. Graetzel, E. Thimsen, *Nature Mater.* **2011**, *10*, 456.
[13] K. Hagedorn, S. Collins, S. Maldonado, *J. Electrochem. Soc.* **2010**, *157*, D588.
[14] O. Khaselev, J. A. Turner, *Science* **1998**, *280*, 425.
[15] B. Marsen, B. Cole, E. L. Miller, *Sol. Energy Mater. Sol. Cells* **2008**, *92*, 1054.
[16] A. Wolcott, W. A. Smith, T. R. Kuykendall, Y. P. Zhao, J. Z. Zhang, *Small* **2009**, *5*, 104.
[17] C. A. Grimes, O. K. Varghese, S. Ranjan, *Light, Water, Hydrogen - The Solar Generation of Hydrogen by Water Photoelectrolysis* Springer, New York, **2008**.
[18] A. Wolcott, W. A. Smith, T. R. Kuykendall, Y. P. Zhao, J. Z. Zhang, *Adv. Funct. Mater.* **2009**, *19*, 1849.